\documentclass[preprint,aps,prc,superscriptaddress,showpacs,floatfix]{revtex4}
\usepackage{graphicx}
\usepackage{dcolumn}
\usepackage{bm}

\begin{document}

\preprint{}
\title{Double neutron-proton differential transverse flow as
a probe for the high-density behavior of the nuclear symmetry
energy}
\author{Gao-Chan Yong}
\affiliation{Institute of Modern Physics, Chinese Academy of Science, Lanzhou 730000,
China}
\affiliation{Graduate School, Chinese Academy of Science, Beijing 100039, P.R. China}
\author{Bao-An Li}
\affiliation{Department of Physics, Texas A\&M University-Commerce, Commerce, TX 75429,
and Department of Chemistry and Physics, P.O. Box 419, Arkansas State
University, State University, AR 72467-0419, USA }
\author{Lie-Wen Chen}
\affiliation{Institute of Theoretical Physics, Shanghai Jiao Tong University, Shanghai
200240, China}
\affiliation{Center of Theoretical Nuclear Physics, National Laboratory of Heavy Ion
Accelerator, Lanzhou 730000, China }

\begin{abstract}
The double neutron-proton differential transverse flow taken from
two reaction systems using different isotopes of the same element
is studied at incident beam energies of $400$ and $800$
MeV/nucleon within the framework of an isospin- and
momentum-dependent hadronic transport model IBUU04. The double
differential flow is found to retain about the same sensitivity to
the density dependence of the nuclear symmetry energy as the
single differential flow in the more neutron-rich reaction.
Because the double differential flow reduces significantly both
the systematic errors and the influence of the Coulomb force, it
is thus more effective probe for the high-density behavior of the
nuclear symmetry energy.
\end{abstract}

\maketitle
\date{\today}

\section{INTRODUCTION}

The density dependence of the nuclear symmetry energy is not only important
for nuclear physics, but also crucial for many astrophysical processes, such
as the structure of neutron stars and the dynamical evolution of
proto-neutron stars \cite{mk94,ba0511}. Heavy-ion reactions induced by
neutron-rich nuclei, especially radioactive beams,
provide a unique opportunity to constrain the equation of
state (EOS) of asymmetric nuclear matter \cite{ireview,ibook,baran05}.
Though considerable progress has been made recently in determining the
density dependence of the nuclear symmetry energy around the normal nuclear
matter density from studying the isospin diffusion in heavy-ion reactions at
intermediate energies \cite{mbt,chen04,li05}, much more work is still needed
to probe the high-density behavior of the nuclear symmetry energy.

A key task is to identify experimental observables sensitive to the density
dependence of the nuclear symmetry energy, especially at high densities.
Several potentially useful observables, such as, the free neutron/proton
ratio \cite{ba97a}, the isospin fractionation \cite%
{serot,ba97b,vb98,hs00,wp01,vb02}, the neutron-proton correlation function
\cite{lw03a}, $t$/$^{3}$He \cite{chen03a,zhang05}, the isospin diffusion
\cite{lw04a,ls03}, the proton differential elliptic flow \cite{ba01a} and
the $\pi ^{-}/\pi ^{+}$ ratio \cite{ba02a,gai04,qli05a,qli05b} have been
proposed in the literature. The concept of the neutron-proton differential
flow was first introduced by one of us \cite{ba00a} several years ago. It
was argued that the neutron-proton differential flow minimizes influences of
the isoscalar potential but maximizes effects of the symmetry potential. It
can also reduce effects of other dynamical ingredients in intermediate
energy heavy-ion reactions. It is therefore among the most promising probes
for the high density behavior of the nuclear symmetry energy.

In order to extract accurately information about the symmetry
energy one has to reduce as much as possible the systematic errors
involved in the experimental observables. Moreover, the long range
Coulomb force on charged particles may play an important role in
these observables. If at all possible, one would like to
disentangle effects of the symmetry energy from those due to the
Coulomb force. Very often, this is impossible. One would thus like
to construct observables that can reduce the Coulomb effects as
much as possible. Ratios and/or differences of two observables
from a pair of reactions using different isotopes of the same
element are among the promising candidates to reduce both the
systematic errors and the Coulomb effects. Whether to use the
ratio or the difference to construct the desired observable
depends on the nature of the observables involved. For the
neutron/proton ratio of pre-equilibrium nucleons and the $\pi
^{-}/\pi ^{+}$ ratio, for instance, it is natural to construct
their double ratios as was recently done in
Refs.\cite{li05b,yong06,msu06}. However, the neutron-proton
differential flow is additive, it is more useful to construct the
double differences instead of ratios. In the present work, we
investigate the double neutron-proton differential transverse flow
from the two reactions of $^{132}$Sn+$^{124}$Sn and
$^{112}$Sn+$^{112}$Sn at beam energies of $400$ and $800$
MeV/nucleon. It is found to have the same sensitivity to the
density dependence of the nuclear symmetry energy as the single
neutron-proton differential flow in the neutron-rich reaction
system $^{132}$Sn+$^{124}$Sn. Besides having smaller systematic
errors, the double differential flow is shown indeed to reduce
significantly the Coulomb effects.

\section{A BRIEF INTRODUCTION TO THE IBUU04 TRANSPORT MODEL}

\begin{figure}[th]
\begin{center}
\includegraphics[width=0.5\textwidth]{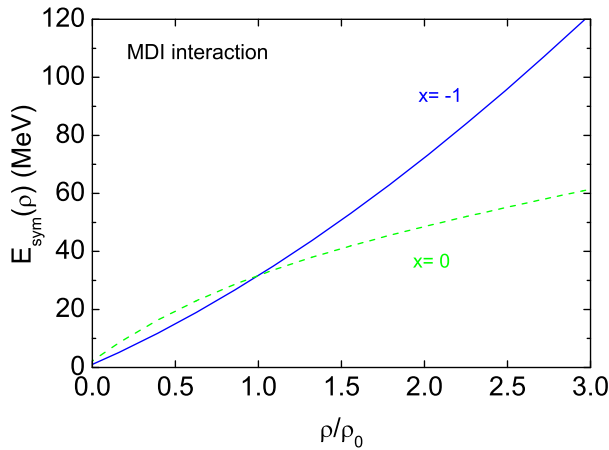}
\end{center}
\caption{{\protect\small (Color online) Density dependence of nuclear
symmetry energy using the MDI interaction with $x=0$ and $x=-1$.}}
\label{sym}
\end{figure}

Our present studies are based on the transport model IBUU04, in which
nucleons, $\Delta $ and $N^{\ast }$ resonances as well as pions and their
isospin-dependent dynamics are included. The initial neutron and proton
density distributions of the projectile and target are obtained by using the
relativistic mean field theory. We use the isospin-dependent in-medium
nucleon-nucleon (NN) elastic cross sections from the scaling model according
to nucleon effective masses \cite{li05}. For the inelastic cross sections we
use the experimental data from free space NN collisions since at higher
incident beam energies the NN cross sections have no evident effects on the
slope of neutron-proton differential flow \cite{li05}. The total and
differential cross sections for all other particles are taken either from
experimental data or obtained by using the detailed balance formula. The
isospin dependent phase-space distribution functions of the particles
involved are solved by using the test-particle method numerically. The
isospin-dependence of Pauli blockings for fermions is also considered.
Details can be found in Refs. \cite{ba97a,ba04a,das03,ba04,li05,yong06}. The
momentum- and isospin-dependent single nucleon potential (MDI) adopted \cite%
{das03} is
\begin{eqnarray}
U(\rho ,\delta ,\mathbf{p},\tau ) &=&A_{u}(x)\frac{\rho _{\tau ^{\prime }}}{%
\rho _{0}}+A_{l}(x)\frac{\rho _{\tau }}{\rho _{0}}  \nonumber \\
&&+B(\frac{\rho }{\rho _{0}})^{\sigma }(1-x\delta ^{2})-8x\tau \frac{B}{%
\sigma +1}\frac{\rho ^{\sigma -1}}{\rho _{0}^{\sigma }}\delta \rho _{\tau
^{\prime }}  \nonumber \\
&&+\frac{2C_{\tau ,\tau }}{\rho _{0}}\int d^{3}\mathbf{p}^{\prime }\frac{%
f_{\tau }(\mathbf{r},\mathbf{p}^{\prime })}{1+(\mathbf{p}-\mathbf{p}^{\prime
})^{2}/\Lambda ^{2}}  \nonumber \\
&&+\frac{2C_{\tau ,\tau ^{\prime }}}{\rho _{0}}\int d^{3}\mathbf{p}^{\prime }%
\frac{f_{\tau ^{\prime }}(\mathbf{r},\mathbf{p}^{\prime })}{1+(\mathbf{p}-%
\mathbf{p}^{\prime })^{2}/\Lambda ^{2}}.  \label{potential}
\end{eqnarray}%
The detailed values of the parameters can be found in Ref. \cite%
{das03,chen04,li05}. With the above single particle potential $U(\rho
,\delta ,\mathbf{p},\tau )$, for a given value $x$, one can readily
calculate the symmetry energy $E_{\text{sym}}(\rho )$ as a function of
density. Noticing that the isospin diffusion data from NSCL/MSU have
constrained the value of $x$ to be between $0$ and $-1$ for nuclear matter
densities less than about $1.2\rho _{0}$ \cite{chen04,li05}, in the present
work, as an example, we also consider the two values of $x=0$ and $x=-1$.
Shown in Fig.\ \ref{sym} is the density dependence of the nuclear symmetry
energy with the two $x$ values. It is seen that the case of $x=0$ gives a
softer symmetry energy than that of $x=-1$ and the difference becomes larger
at higher densities.

\section{RESULTS AND DISCUSSIONS}

\begin{figure}[th]
\begin{center}
\includegraphics[width=0.5\textwidth]{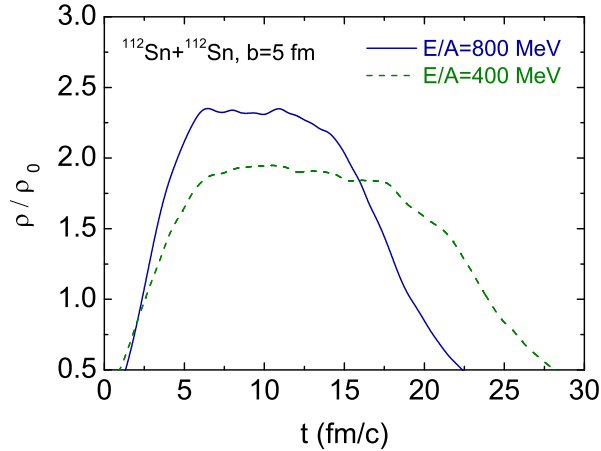}
\end{center}
\caption{{\protect\small (Color online) Time evolution of the average baryon
density in the central cell reached in the $^{112}$Sn+$^{112}$Sn reaction at
the incident beam energies of $400$ and $800$ MeV/nucleon with an impact
parameter of }${\protect\small b=5}${\protect\small \ fm.}}
\label{density}
\end{figure}
To study the high-density behavior of the symmetry energy, it is useful to
know the maximal baryon density reached in a given reaction. The maximal
density depends not only on the incident beam energy, but also on the impact
parameter as well as the reaction system. Fig.\ \ref{density} shows the time
evolution of the average baryon density in the central cell reached in $%
^{112}$Sn+$^{112}$Sn reaction at beam energies of $400$ and $800$
MeV/nucleon with an impact parameter of $b=5$ fm. One can see that for $400$
MeV/nucleon, the maximal baryon density reached is about $1.9\rho _{0}$
while it is about $2.4\rho _{0}$ for $800$ MeV/nucleon. One can also see
that at the higher incident energy, the life time of the high density
nuclear matter is shorter as expected.

The neutron-proton differential transverse flow was defined as \cite{ba00a,
ba02a}
\begin{eqnarray}
F_{n-p}^{x}(y) &\equiv &\frac{1}{N(y)}\sum_{i=1}^{N(y)}p_{i}^{x}(y)w_{i}
\nonumber \\
&=&\frac{N_{n}(y)}{N(y)}\langle p_{n}^{x}(y)\rangle -\frac{N_{p}(y)}{N(y)}%
\langle p_{p}^{x}(y)\rangle  \label{npflow}
\end{eqnarray}%
where $N(y)$, $N_{n}(y)$ and $N_{p}(y)$ are the number of free
nucleons, neutrons and protons, respectively, at rapidity $y$;
$p_{i}^{x}(y)$ is the transverse momentum of the free nucleon at
rapidity $y$; $w_{i}=1$ $(-1)$ for neutrons (protons); and
$\langle p_{n}^{x}(y)\rangle $ and $\langle p_{p}^{x}(y)\rangle $
are respectively the average transverse momenta of neutrons and
protons at rapidity $y$. It is seen from Eq. (\ref{npflow}) that
the constructed neutron-proton differential transverse flow
depends not only on the proton and neutron transverse momenta but
also on their relative multiplicities. We stress that the
neutron-proton differential flow combines effects due to both the
isospin fractionation and the different transverse flows of
neutrons and protons. It is noticed that the neutron-proton
differential transverse flow is not simply the difference of the
neutron and proton transverse flows. Instead, it depends also on
the isospin fractionation at the rapidity $y$. To see this point
more clearly, let's consider two special cases. If neutrons and
protons have the same average transverse momentum in the reaction
plane but different multiplicities in each rapidity bin, i.e.,
$\langle p_{n}^{x}(y)\rangle =\langle p_{p}^{x}(y)\rangle =\langle
p^{x}(y)\rangle $, and $N_{n}(y)\neq N_{p}(y)$, then Eq.
(\ref{npflow}) is reduced to
\begin{equation}
F_{n-p}^{x}(y)=\frac{N_{n}(y)-N_{p}(y)}{N(y)}\langle p^{x}(y)\rangle =\delta
(y)\cdot \langle p^{x}(y)\rangle,
\end{equation}%
reflecting effects of the isospin fractionation. On the other hand, if
neutrons and protons have the same multiplicity but different average
transverse momenta, i.e., $N_{n}(y)=N_{p}(y)$ but $\langle
p_{n}^{x}(y)\rangle \neq \langle p_{p}^{x}(y)\rangle $, then Eq. (\ref%
{npflow}) is reduced to%
\begin{equation}
F_{n-p}^{x}(y)=\frac{1}{2}(\langle p_{n}^{x}(y)\rangle -\langle
p_{p}^{x}(y)\rangle ).
\end{equation}%
In this case it reflects directly the difference of the neutron and proton
transverse flows. In heavy-ion collisions at higher energies \cite%
{li05b,yong06}, generally, for free nucleons in a given rapidity bin, one
expects that a stiffer symmetry potential leads to a higher isospin
fractionation and also contributes more positively to the transverse momenta
of neutrons compared to protons. The neutron-proton differential flow thus
combines constructively effects of the symmetry potentials for neutrons and
protons.
\begin{figure}[th]
\begin{center}
\includegraphics[width=0.5\textwidth]{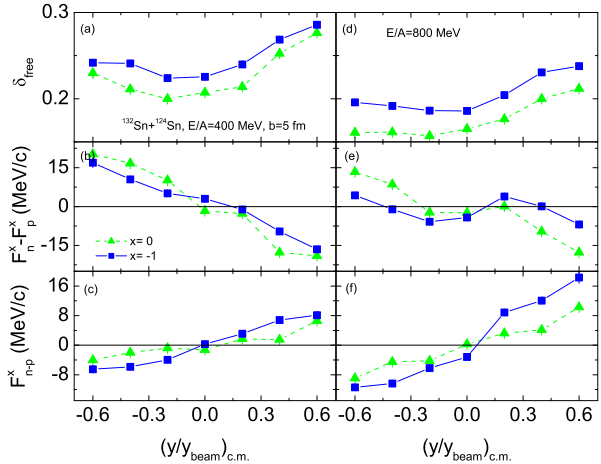}
\end{center}
\caption{(Color online)  Rapidity distribution of the isospin
asymmetry of free nucleons (upper panels), the difference of the
average nucleon transverse flows (middle panels) and the
neutron-proton
differential transverse flow (lower panels) from $^{132}$%
Sn+$^{124}$Sn reaction at the incident beam energies of $400$,
$800$ MeV/nucleon and $b=5$ fm with two symmetry energies of $x=0$
and $x=-1$.} \label{isoflow1}
\end{figure}

Shown in Fig.\ \ref{isoflow1} are the rapidity distribution of the
isospin asymmetry of free nucleons (upper panels), the difference
of the average nucleon transverse flows (middle panels) and the
neutron-proton differential transverse flow (lower panels) from
the $^{132}$Sn+$^{124}$Sn reaction at incident beam energies of
$400$, $800$ MeV/nucleon and an impact parameter of $b=5$ fm with
the two symmetry energies of $x=0$ and $x=-1$. It is seen from the
upper panels of Fig.\ \ref{isoflow1} that a larger isospin
asymmetry of free nucleons (stronger isospin fractionation) is
obtained for the stiffer symmetry energy ($x=-1$). It is
interesting to see from the bottom panels of Fig.\ \ref{isoflow1}
that the stiffer symmetry energy ($x=-1$) leads to clearly a
stronger neutron-proton differential transverse flow than the
softer symmetry energy ($x=0$). From the middle panels we notice
that the difference of the average nucleon flows exhibits less
sensitivity to the symmetry energy compared with the
neutron-proton differential transverse flow. Normally, the Coulomb
potential dominates over the symmetry potential for protons,
consequently protons have higher average transverse momenta than
neutrons, leading to the negative (positive) values of the
$F_n^x-F_p^x$ at forward (backward) rapidities.

We examine the beam energy dependence of the neutron-proton
differential transverse flow in the lowest two panels (c) and (f)
of Fig.\ \ref{isoflow1}. As one expects, with the same symmetry
energy, the slope of the neutron-proton differential transverse
flow around the mid-rapidity is larger for the higher incident
beam energy. This is mainly because a denser nuclear matter is
formed at higher incident beam energy (shown in Fig.\
\ref{density}). It then leads to a stronger symmetry potential and
thus higher transverse momenta for neutrons compared to protons.
The magnitude of the neutron-proton differential transverse flow
at $800$ MeV/nucleon is much larger than that at $400$ MeV/nucleon
and it is thus easier to be measured experimentally although the
net effect of the symmetry potential on the neutron-proton
differential transverse flow is not much larger than that at $400$
MeV/nucleon.
\begin{figure}[th]
\begin{center}
\includegraphics[width=0.5\textwidth]{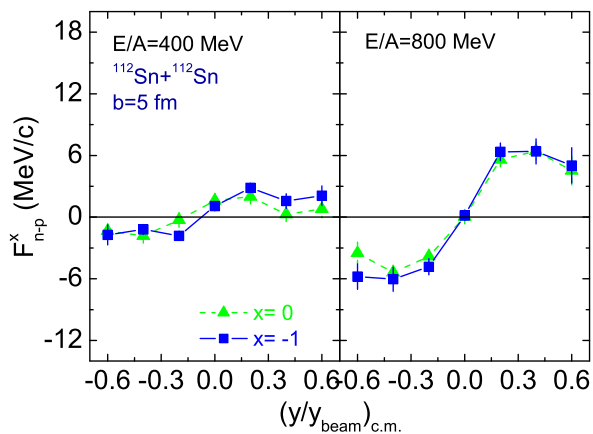}
\end{center}
\caption{{\protect\small (Color online) Same as the lowest two
panels (c) and (f) of Fig.\ \protect\ref{isoflow1} but for the
reaction system of $^{112}$Sn+$^{112}$Sn. }} \label{flow112}
\end{figure}

In order to reduce the systematic errors, one can study the relative values
of some observables from two similar reaction systems. In the present work,
we thus also studied the less neutron-rich reaction system $^{112}$Sn+$%
^{112} $Sn with the same reaction conditions as a reference. Fig.\ \ref%
{flow112} shows the rapidity distribution of the neutron-proton differential
transverse flow in the semi-central reaction of $^{112}$Sn+$^{112}$Sn at the
same incident beam energies of $400$ and $800$ MeV/nucleon. Comparing with
the case of $^{132}$Sn+$^{124}$Sn, we can see that the slope of the
neutron-proton differential transverse flow around mid-rapidity and effects
of the symmetry energy become much smaller due to the smaller isospin
asymmetry in the reaction of $^{112}$Sn+$^{112}$Sn.
\begin{figure}[th]
\begin{center}
\includegraphics[width=0.5\textwidth]{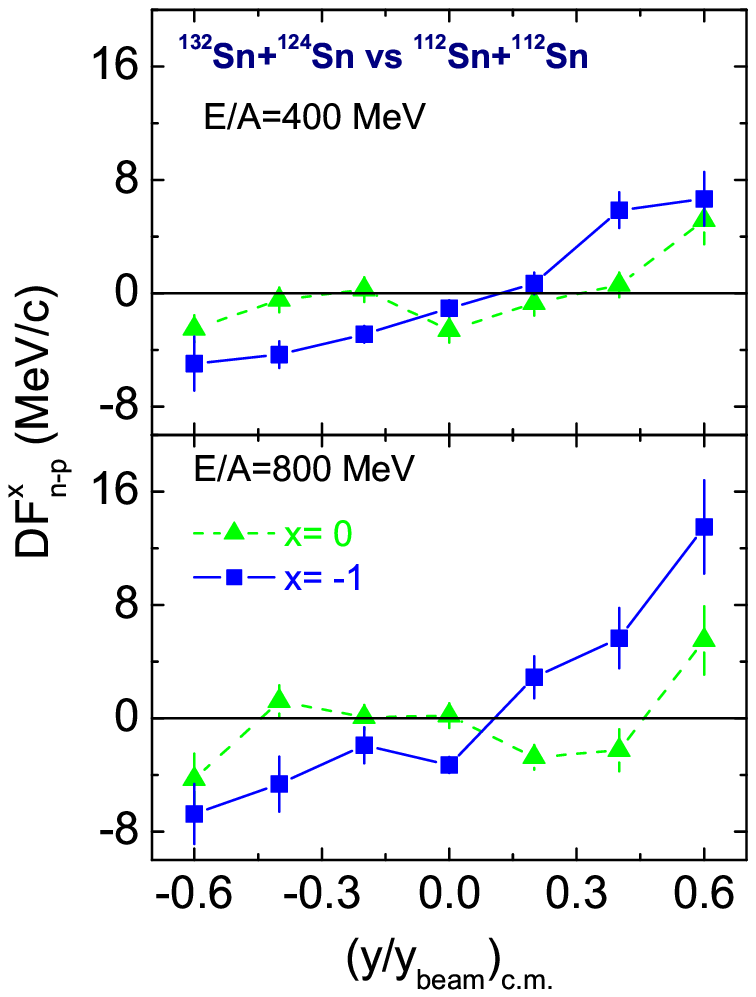}
\end{center}
\caption{{\protect\small (Color online) Rapidity distribution of the double
neutron-proton differential transverse flow in the semi-central reactions of
Sn+Sn isotopes at the incident beam energies of $400$ and $800$ MeV/nucleon
with two symmetry energies of $x=0$ and $x=-1$.}}
\label{dflow}
\end{figure}
\begin{figure}[th]
\begin{center}
\includegraphics[width=0.5\textwidth]{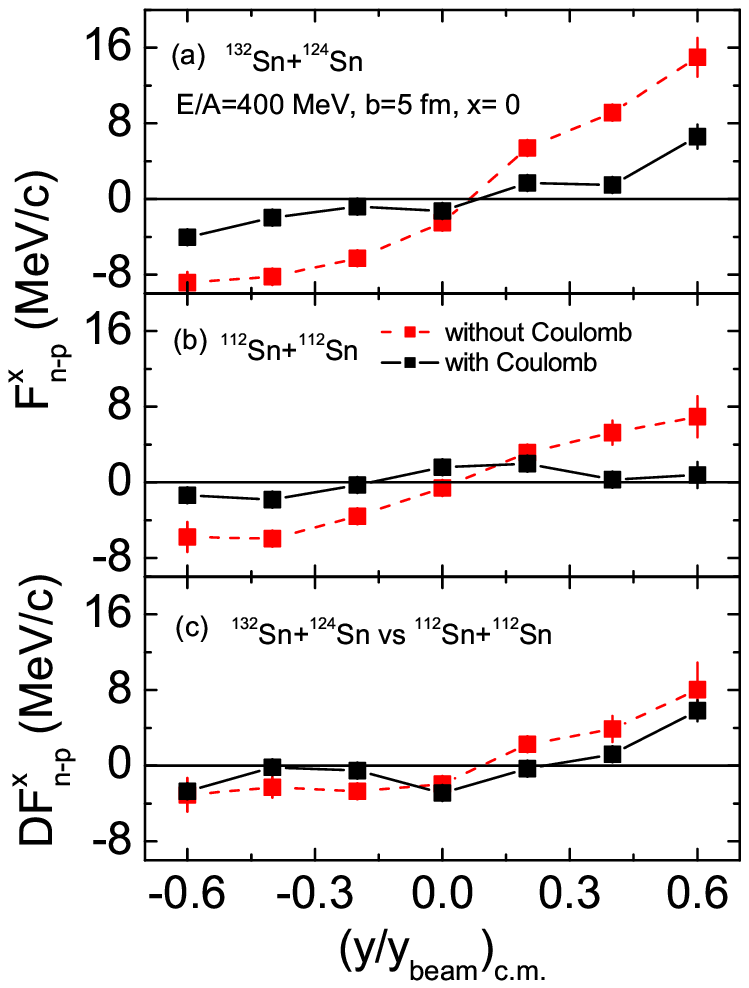}
\end{center}
\caption{{\protect\small (Color online) Coulomb effects on the
neutron-proton differential transverse flow (upper two panels) and the
double neutron-proton differential transverse flow (lowest panel) in the
semi-central reactions of Sn+Sn isotopes at the incident beam energy of $400$
MeV/nucleon with the symmetry energy of $x=0$.}}
\label{Coulomb}
\end{figure}

How should one use the reaction of $^{112}$Sn+$^{112}$Sn as a
reference? Since for $^{112}$Sn+$^{112}$Sn effects of the symmetry
energy on neutron-proton differential transverse flow almost
disappear, it is the easiest to study the difference of the
neutron-proton differential flows (we dub it the double
neutron-proton differential flow) from the two reaction systems of
$^{132}$Sn+$^{124}$Sn and $^{112}$Sn+$^{112}$Sn. Fig.\ \ref{dflow}
shows the rapidity distribution of the double neutron-proton
differential transverse flow in the semi-central reactions of
Sn+Sn isotopes. At both incident beam energies of $400$ and $800$
MeV/nucleon, it is interesting to see that the double
neutron-proton differential transverse flow around mid-rapidity is
essentially zero for the soft symmetry energy of $x=0$. However,
it displays a clear slope with respect to the rapidity for the
stiffer symmetry energy of $x=-1$. Moreover, the double
neutron-proton differential transverse flow at the higher incident
energy indeed exhibits a stronger symmetry energy effect as
expected. Furthermore, it is seen that the double neutron-proton
differential transverse flow retains about the same symmetry
energy effect as the $^{132}$Sn+$^{124}$Sn reaction. As discussed
in Ref. \cite{rami00}, observables coming from many reaction
combinations under identical experimental conditions are
insensitive to systematic uncertainties due to the apparatus.
Theoretically, in transport model calculations, the systematic
errors are mostly related to the physical uncertainties of
in-medium NN cross sections, techniques of treating collisions,
sizes of the lattices in calculating the phase space
distributions, techniques in handling the Pauli blocking, etc..
Since the double neutron-proton differential flow is a relative
observable from the two similar reaction systems, systematic
errors are thus expected to be reduced.

Moreover, the Coulomb force on charged particles may play important roles.
It sometimes competes strongly with the symmetry potentials.
One thus has to disentangle carefully effects of
the symmetry potentials from those due to the Coulomb potentials. Because
the double neutron-proton differential transverse flow is a relative
observable, Coulomb effects are also expected to be much reduced.
To verify this expectation, we examine in Fig.\ \ref{Coulomb} Coulomb effects
on the neutron-proton differential transverse flow (upper two panels)
and the double neutron-proton differential transverse flow (lowest panel)
in the semi-central reactions of Sn+Sn isotopes at the incident
beam energy of $400$ MeV/nucleon with the symmetry energy of $x=0$.
From the upper two panels of Fig.\ \ref{Coulomb},
one can see that the Coulomb effects reduce the strength of the
neutron-proton differential transverse flow. With the Coulomb force, more
protons are unbound and have large transverse momenta in the reaction-plane.
According to Eq. (\ref{npflow}), the strength of the neutron-proton
differential transverse flow will be reduced. The lowest panel of Fig.\ \ref%
{Coulomb} shows that the double neutron-proton differential transverse flow
can reduce the effect of long range Coulomb force largely.

\section{SUMMARY}

In summary, based on the isospin- and momentum-dependent hadronic
transport model IBUU04, we have studied the single and double
neutron-proton differential transverse flow and its dependence on
the nuclear symmetry energy in the semi-central reactions of $^{132}$Sn+$^{124}$%
Sn and $^{112}$Sn+$^{112}$Sn at beam energies of $400$ and $800$
MeV/nucleon. We find that the double neutron-proton differential
flow retains about the same sensitivity to the symmetry energy as
the single differential flow in the more neutron-rich system
involved. Because the double neutron-proton differential flow can
reduce significantly both the systematic errors and effects of the
Coulomb force, it is thus more useful probe for the high-density
behavior of the nuclear symmetry energy.

\section*{ACKNOWLEDGMENTS}

We would like to thank Wolfgang Trautmann for his encouragement, continued
interest in this project and helpful discussions. The work was supported in
part by the US National Science Foundation under Grant No. PHY-0354572,
PHY0456890 and the NASA-Arkansas Space Grants Consortium Award ASU15154, the
National Natural Science Foundation of China under Grant No. 10575071, MOE
of China under project NCET-05-0392, and Shanghai Rising-Star Program under
Grant No. 06QA14024.

\end{document}